# Substrate Effects on the Bandwidth of CdSe Quantum Dot Photodetectors

*Christine Schedel[1], Fabian Strauß[1], Krishan Kumar[1], Andre Maier[1], Kai M. Wurst[1], Patrick Michel[1], Marcus Scheele[1,2,*]*

[1]Institute for Physical and Theoretical Chemistry, University of Tübingen, 72076 Tübingen, Germany.

[2]Center for Light-Matter Interaction, Sensors and Analytics LISA+, University of Tübingen, 72076 Tübingen, Germany

We investigate the time-resolved photocurrent response of CdSe quantum dot (QD) thin films sensitized with zinc β-tetraaminophthalocyanine (Zn4APc)[1] on three different substrates, namely silicon with 230 nm $SiO_2$ dielectric, glass as well as polyimide. While Si/$SiO_2$(230 nm) is not suitable for any transient photocurrent characterization due to an interfering photocurrent response of the buried silicon, we find that polyimide substrates invoke the larger optical bandwidth with 85 kHz vs. 67 kHz for the same quantum dot thin film on glass. Upon evaluation of the transient photocurrent, we find that the photoresponse of the CdSe quantum dot films can be described as a combination of carrier recombination and fast trapping within 2.7 ns, followed by slower multiple trapping events. The latter are less pronounced on polyimide, which leads to the higher bandwidth.

We show that all devices are RC-time limited and that improvements of the photoresistance are the key to further increasing the bandwidth.

**INTRODUCTION**

In order to accelerate the speed of a photodetector, most studies focus on the photoabsorber[2-4], the type of electrodes[5-7] and their geometries[8-10]. Due to their large photoabsorption cross-sections, solution-processability, and compatibility with flexible substrates without the restriction of epitaxial growth, colloidal quantum dots are promising candidates for application in photodetectors.[11, 12] Such quantum dot photodetectors are fabricated on many different substrates, such as Si/SiO$_2$[13-15], including our own first experiments[1], glass[16-21], sapphire[22, 23] and flexible polymer foil like polyethylene terephthalate[23-26]. However, very few studies focus on the influence of the chosen substrate on the performance of the photodetector. Few studies have noted that the substrate can have an impact on the responsivity[27, 28] or even on the rise/fall time constants under square pulse illumination[28]. In particular, we are not aware of any study that examines the substrate effect on the bandwidth of otherwise identical photodetectors. Sometimes the substrate is not even mentioned[2, 29] what shows how little attention has been paid to this aspect in investigations to date.

For a full investigation of the photocurrent response, both steady-state and non-steady state properties must be examined. Therefore, we probed the response of CdSe/I$^-$/Zn4APc thin films[1] on Si/SiO$_2$, glass and polyimide substrates under both 635 nm square pulse illumination, reaching the steady-state photoresponse, and 636 nm and 779 nm impulse illumination. We show that silicon with a thin SiO$_2$ dielectric layer is not suitable as substrate material for time-resolved UV/Vis photocurrent measurements due to a strong photocurrent signal of the silicon. While glass or polyimide substrates are feasible alternatives, we find distinctly different 3 dB bandwidths for the same absorber material on the two substrates. On polyimide, a thin film of CdSe/I$^-$/Zn4APc as active material exhibits a bandwidth of 85 kHz, while for the same photodetectors on glass the bandwidth is only 67 kHz. Two photocurrent decay mechanisms can be identified in non-steady state studies: First, a rapid exponential decay that is identical on both substrates, followed by a long trap-state induced tail that is significantly enhanced on glass. The latter is the main cause of the bandwidth difference. In contrast, the decay behavior in steady-state studies shows no difference on either substrate material, indicating that all trap states are saturated due to the long illumination time, leading to comparable decay behavior for all detectors. Moreover, we find that all detectors are RC limited due to relatively high resistances, even for channel lengths of only 350 nm. Thus, for faster quantum dot photodetectors, the photoresistance must be reduced further.

**RESULTS**

**Photoresponse on Si/SiO$_2$ substrates**. In Figure 1, we show why substrates with a silicon bottom contact and a CMOS-quality SiO$_2$ dielectric layer of typical thickness (here: 230 nm) with interdigitated gold top contacts (cf. Figure SI1a) are unsuitable as substrates for lateral phototransistors. We observe a significant photocurrent signal under 636 nm illumination without any photoactive material deposited in the channel, a reduced signal for 779 nm illumination and no signal under 1310 nm laser illumination, suggesting that enough light penetrates the dielectric to excite the silicon below. Consistent with this view, no photocurrent can be measured with a SiO$_2$ dielectric of 770 nm thickness. We therefore advise caution when using Si/SiO$_2$ substrates for lateral phototransistors, since the rise time of the material of interest in the channel is likely to be convoluted with the rise time of the silicon bottom contact. In view of the popularity of these substrates, our findings in Figure 1 may be of relevance for the reported rise times in many studies in the past, including our own results on CdSe-based phototransistors.[1] To avoid any convolution with the rise time of a buried bottom contact, in the present work we investigate the time response of photodetectors fabricated on plain glass and flexible polyimide substrates with Au top contacts, which were verified to exhibit no photocurrent signal when empty.

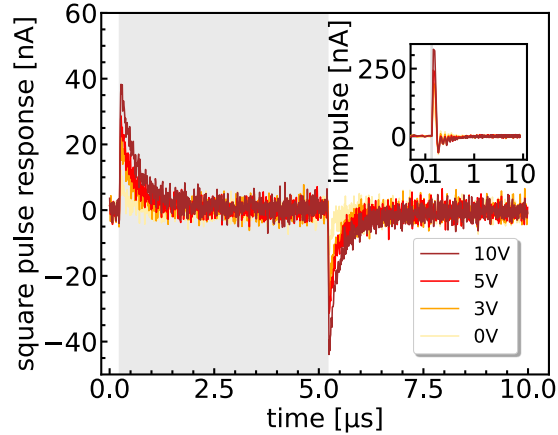

Figure 1. Typical Si-substrate signal for a transistor consisting of a Si/SiO$_2$(230 nm) substrate and interdigitated Au electrodes on top (here: 2.5 μm × 10 mm) under 100 kHz illumination without any active material deposited in the channel. Grey boxes indicate laser illumination, using a 635 nm square pulse laser and a 636 nm impulse laser, shown in the inset.

**Non-steady state (or "impulse") photoresponse on glass and polyimide substrates.** Transient photocurrent response data are of particular interest for the possible use as photodetectors in fast optical switches, since very short, delta-shaped laser impulses are used in optical communication. A schematic of the transient photoconductivity measurement setup is shown in Figure SI2. In a comparison of CdSe/I$^-$/Zn4APc thin films on glass and polyimide foil with μm-spaced electrodes (see Figure SI1), an important difference can be observed in the data regarding the speed of response of the devices, cf. Figure 2.

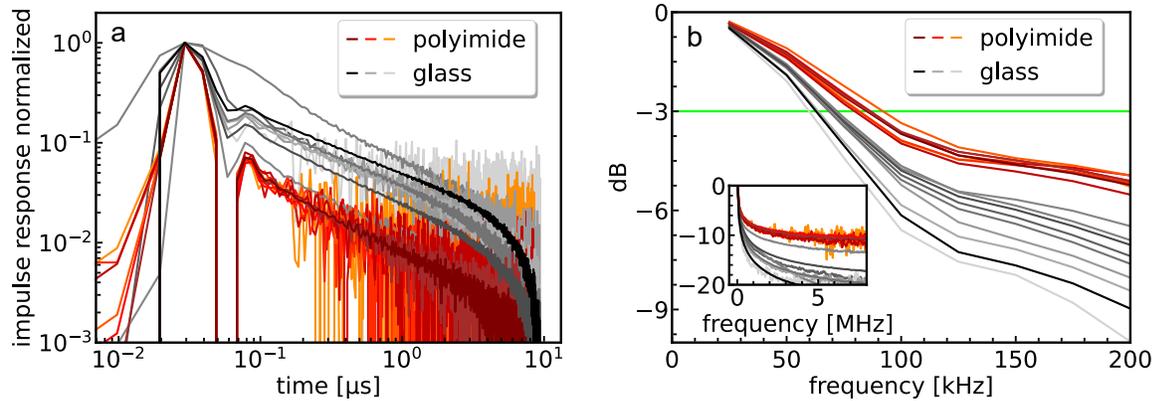

Figure 2. a) Normalized impulse responses of several CdSe/I$^-$/Zn4APc thin films on polyimide (red) and glass (grey) for 10 V, 636 nm 100 kHz measurements (corresponding to 9.8 ns time/point resolution in these measurements) with µm gap channels. The pronounced, slowly decaying tail in measurements on glass substrates is visible. b) Corresponding bandwidth of these impulse spectra, showing the effect of the tail in the frequency range. Inset: wider range of the bandwidth spectrum.

The normalized impulse photoresponses of CdSe/I$^-$/Zn4APc thin films on µm-spaced electrodes on both glass and polyimide substrates towards a 636 nm laser are shown in Figure 2a. The long tail is more pronounced on glass substrates with a fall time (90 % – 10 %) of 18 $\pm$ 3 ns on polyimide vs. 160 $\pm$ 128 ns on glass. This difference results in 3 dB bandwidths for CdSe/I$^-$/Zn4APc thin films of 85 $\pm$ 4 kHz on polyimide vs. 67 $\pm$ 4 kHz on glass, cf. Figure 2b (for details of the bandwidth calculation, see section Experimental Methods – Transient Photocurrent). The FWHM is identical on both substrates with approx. 4 ns for both, 636 nm and 779 nm illumination, (evaluated from measurements with higher time resolution and thus shorter measurement windows, not shown here) and there is neither a field- nor a channel length dependence of the FWHM, cf. Figure SI3.

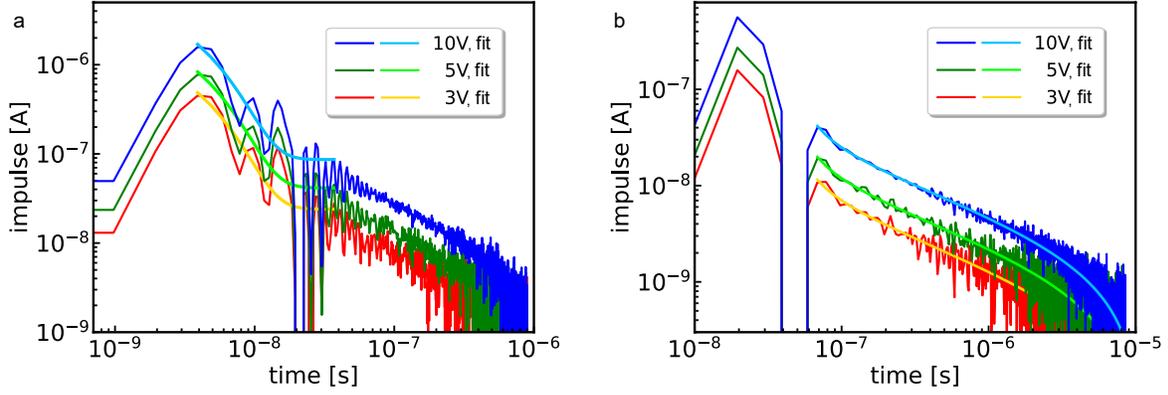

Figure 3. Typical polyimide device impulse response at 636 nm of a 2.5 μm contact with a) exponential fits ($\propto e^{-\frac{t}{2.7\,ns}}$) for the initial decay using 1 MHz repetition rate data, b) power law fits ($\propto t^m$, $m_{10V}$ = -0.47±0.01, $m_{5V}$ = -0.47±0.02, $m_{3V}$ = -0.49±0.03) using 100 kHz repetition rate data for better visualization of the long tail.

We use the time-resolved photoresponse to gain insights into possible decay mechanisms of the excited state in the photodetectors and apply two models: 1) A fast exponential decay ($\propto e^{-\frac{t}{[ns]}}$) on short time scales, which we attribute to carrier recombination and initial trapping[26] and 2) a slower decay following a power-law ($\propto t^m$) according to a multiple trapping and release mechanism.[30] The latter model assumes that all trap states have equal capture cross sections, which invokes a shift in the population density of the trap states toward deeper traps with time, i.e. after several trap and release steps, as charge carriers in shallow traps are preferentially released. Thereby, an exponentially decreasing trap state density towards deeper traps results in the above-mentioned power law decay. The trap state distribution is expressed with the Urbach energy $E_U$, which can be calculated via the exponent $m$ of the power law fit ($\propto t^m$) using $E_U = \frac{k_B T}{m+1}$.[26] We observe both mechanisms in our data, similar to other reports.[25, 26] Initially, an exponential decay is visible, which transforms into a power law over time. Exemplary fits for both regimes can be seen in Figure 3.

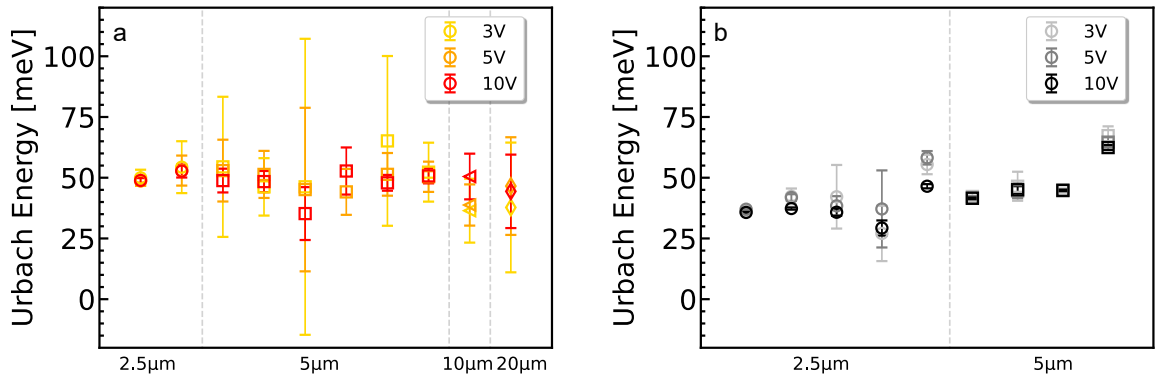

Figure 4. Urbach energy of the investigated devices for a) polyimide (red) and b) glass (grey) substrates, x-axis: electrode gap (µm × 10 mm), contacts listed.

The time constant calculated via the exponential fit of the first signal decay is about 2.7 ns for both substrate materials regardless of the applied electric field and channel length. The mean of all Urbach energies of the polyimide devices is approx. 50 meV, while for glass photodetectors it is approx. 44 meV, cf. Figure 4. There seems to be a trend toward decreasing Urbach energy with increasing applied voltage, although most differences are within the standard deviation. The smaller standard deviation for glass substrates can be explained by the enhanced tail currents resulting in improved signal/noise ratios.

**Steady state (or "square pulse") photoresponse on glass and polyimide substrates**. The photoresponse of a CdSe/I⁻/Zn4APc thin film with µm electrode gap on a polyimide substrate towards a 635 nm laser at 10 Hz square pulse illumination (≤ 12 mW) is shown in Figure 5. The same stable photocurrent is also measurable for longer pulses, such as 0.1 Hz pulses, confirming that steady state conditions are reached. Neither a field dependence nor a channel length dependence of both rise and fall times (10 – 90 %) can be observed which is an indication of a RC limitation of the devices. Likewise, no difference between the two types of substrates can be found, contrary to E *et al.*[28] Rise times vary mostly between 0.4 ms and 4.8 ms (Figure 6b) and fall times between 0.2 ms and 4.2 ms. The scattering can be explained by inhomogeneities in morphology and thickness of the films covering the electrodes (Figure SI1b).

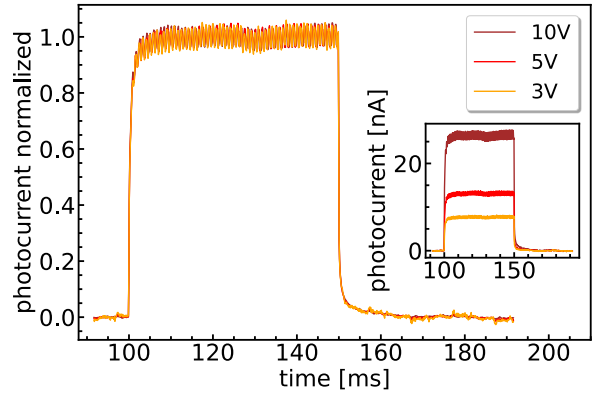

Figure 5. Typical normalized photoresponse of CdSe/I$^-$/Zn4APc on polyimide towards a 635 nm 10 Hz square pulsed laser at different voltages (channel length: 5 µm). Inset: absolute photoresponse.

To test for a possible RC limitation, we performed impedance spectroscopy on the devices on polyimide. An example impedance spectrum can be seen in Figure SI4, including the performed fit for a parallel circuit consisting of a resistor and a constant phase element, similar to Livache *et al.*[18] The resistances from impedance spectroscopy agree well with those from two-point measurements (Figure SI5). As detailed in Figure 6a, the experimental capacitances $C$ are consistent with the theoretically expected values according to $C = L(N-1)\varepsilon_0(1+\varepsilon_r)\frac{K(k)}{K(k')}$,[31] with the finger length ($L$), the number of electrode fingers ($N$), the vacuum permittivity ($\varepsilon_0$), the dielectric constant of CdSe QDs ($\varepsilon_r$, here taken as 6.2)[32], the complete elliptical integral of first kind ($K$) with $k' = \sqrt{1-k^2}$ and $k = \cos\left(\frac{\pi}{2}\left(1 - \frac{w}{w+g}\right)\right)$, the width of the electrode fingers ($w$), and the channel length ($g$). The resulting RC time constants are given in Figure 6b. We note that for all applied voltages, the rise times (10 % - 90 %) are several orders of magnitude faster than the RC times evaluated by impedance spectroscopy in the dark. On the contrary, we obtain excellent agreement between RC and rise times when calculating the RC time constant with the photoresistance ($R_{illum}$).

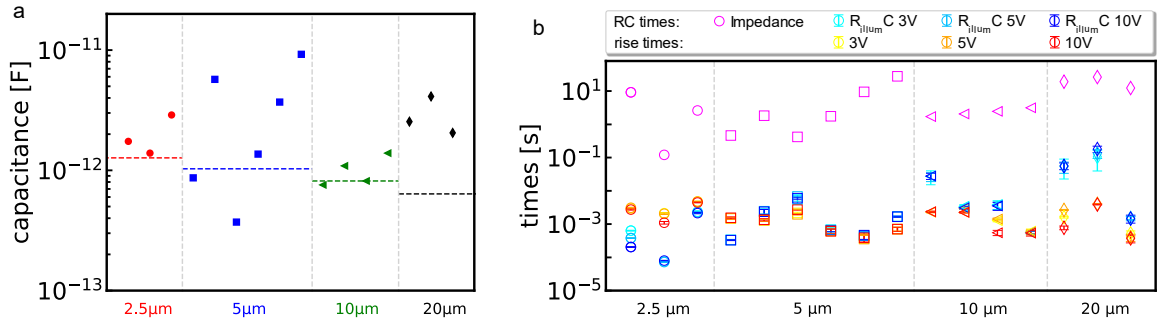

Figure 6. Data of polyimide devices: a) Capacitances for the four different investigated channel lengths. Dashed lines are theoretically calculated capacitances according to Nabet et al.[31] b) Time constants of the devices. Magenta: RC time of impedance measurements in the dark, blue: RC times of the same devices under illumination for different applied electric fields, red: corresponding rise times. x-axis: electrode gap (µm × 10 mm), contacts listed. The three measured values with significantly higher $R_{illum}C$ time compared to the rise time can be considered outliers due to the very low photocurrent levels and associated difficulties in evaluating the data in these devices.

**Reduction of RC time.** In order to reduce the RC time, we now attempt decreasing the resistance in the devices by narrowing the channel length (cf. Figure SI5). To this end we use a new device geometry with shorter channel lengths of 350 nm and 500 nm (channel width 10 mm, interdigitated geometry, see Figure SI6) on glass substrates and compare this third photodetector category (named "nm glass devices") with the first two categories discussed so far (from now on referred to as "µm polyimide" and "µm glass" photodetectors, respectively). Under steady-state conditions, faster rise times down to 24 µs can indeed be achieved with the nm glass photodetectors due to a reduced photoresistance. The relationship between the magnitude of the photoresistance and rise time is depicted in Figure 7. Figure 7a shows the effect of varying photoresistances for both, µm (black) and nm (blue), glass devices, revealing a trend to lower photoresistances and faster rise times for the nm detectors. Identical photoresistances lead to the same rise times for both geometries. This indicates that the devices are still RC limited, which is confirmed by a lack of field dependence of the rise times of the nm-devices (analogous to Figure 5) and the calculation of expected RC times via impedance data (analogous to Figure 6b). Figure 7b shows a typical example for the photoresponses of one CdSe/I$^-$/Zn4APc thin film with different photocurrent heights induced by a 635 nm laser at 100 Hz square pulse illumination, using the 350 nm electrode geometry on glass. A higher photocurrent, that is, a lower photoresistance correlates with a faster detector response under otherwise identical conditions.

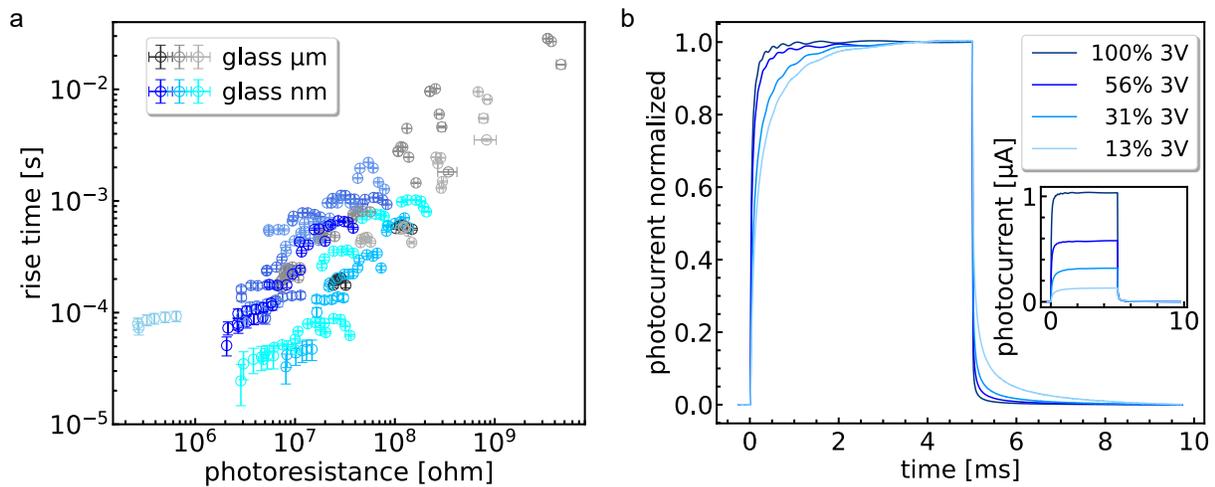

Figure 7. a) Correlation between rise times vs. photoresistance for the μm and nm glass devices (one color = one device). b) Normalized photoresponse of a CdSe/I$^-$/Zn4APc detector on glass towards a 635 nm 100 Hz square pulsed laser with different laser intensities. Inset: absolute photoresponse. Example of a 350 nm channel.

Identical to the μm devices, we use the non-steady state photoresponse to evaluate the bandwidth of the nm photodetectors and plot the results for all three device types in Figure 8a.

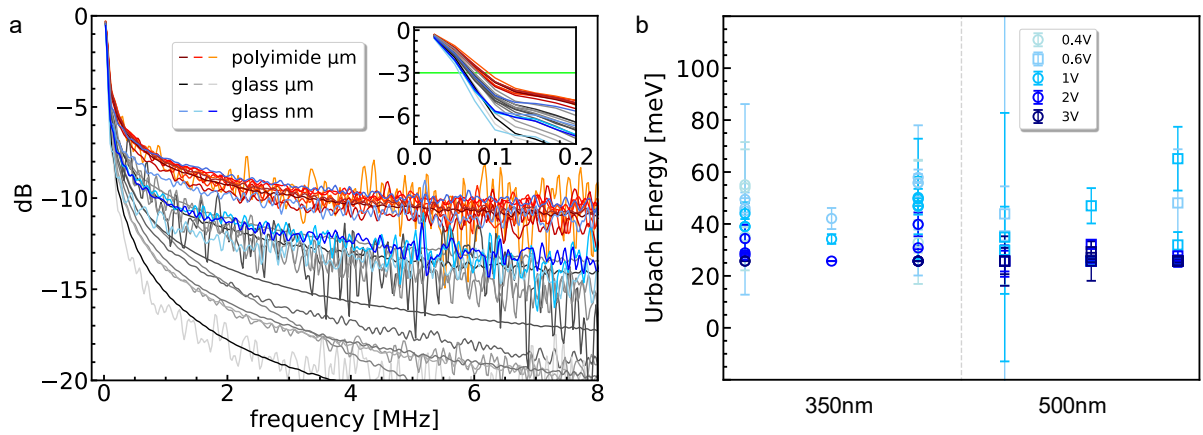

Figure 8. a) Bandwidth spectrum of the nm glass, μm glass, and μm polyimide devices. The corresponding time domain data is given in Figure SI7. Inset: enlarged view on the 3 dB bandwidth. b) Urbach energies of the investigated nm glass devices, x-axis: electrode gap (nm × 10 mm), contacts listed.

The bandwidth spectrum of the CdSe/I$^-$/Zn4APc films on nm glass substrates is less attenuated than on μm glass devices and only slightly stronger than on the μm polyimide devices at higher frequencies, cf. Figure 8a. This shows the improvement of the detectors on glass due to the shorter channel, which manifests in reduced photocurrent tailing, cf. Figure SI7. The 3 dB bandwidth is roughly the same for nm and μm glass devices.

As for all μm devices, an initial exponential decay can also be identified in the impulse photoresponse of the nm detectors, which transforms into a power law behavior in the tail. With 2.5 ns ($\propto e^{-\frac{t}{2.5\text{ns}}}$), the time constant calculated via the exponential fit of the first signal decay is the same as for all μm devices, independent of the channel length and applied field. If the same field strengths are applied to the nm photodetectors as to the μm devices, i.e., up to 40 kV/cm, the same Urbach energies of approx. 40-50 meV are observed, cf. Figure 8b. For higher fields (up to 86 kV/cm) the Urbach energies decrease to about 25 meV.

**DISCUSSION**

We examined the non-steady state and steady-state photocurrent response of CdSe/I$^-$/Zn4APc devices. Under impulse illumination, we obtain important characteristics of the devices for application as fast optical switches. On the other hand, the rise time to reach steady-state conditions is often used to characterize a photodetector towards a potential geometry-related speed limitation.

**Non-steady State.** We attribute the faster response of the quantum dot thin film on polyimide (with μm gap electrodes) compared to glass (both μm and nm gap electrodes) to a lower density of trap states. This is evidenced by the fact that the main distinction we find between the three device categories (μm polyimide, μm glass, nm glass) are differences in the tail of the impulse response (Figure 8a, SI7). The tail analysis in Figure 3 allows gauging the depth of the trap states via their Urbach energies (Figure 4, Figure8b). The values of approx. 50 meV on polyimide and around 44 meV on glass devices are typical for amorphous semiconductors and fit to previously reported values of around 20 – 80 meV for CdSe QD thin films.[33] Taking into account the standard deviation, the Urbach energies determined by us are about the same on both substrates, so this cannot explain the obvious differences in tail currents. We therefore believe that there are more trap states of similar energy available in glass devices than on polyimide resulting in a significantly enhanced multiple trapping and release mechanism. This increased accessible number of trap states arises most probably directly at the interface between QDs and substrate material. Technical glass with a relative permittivity of 4.5-8[34] can shield trapped charge carriers more effectively than polyimide with a lower relative permittivity of 3.7[35], allowing more QD trap states near the interface to be occupied on glass. The reduced tail, i.e., the less attenuated bandwidth spectrum, on nm glass compared to µm glass can thus be explained by the reduced active area to about 10 %, resulting in the presence of significantly fewer trap states per contact.

For most of the samples, the Urbach energy decreases with the applied voltage (Figure 4, Figure 8b). The higher the applied field, the easier it is to excite the trapped charge carriers from the trap states into the band which results in an occupation density of the trap states closer to the band edge than for smaller electric fields.

In contrast, the time constant of the very first exponential signal decay is with about 2.7 ns identical for all photodetectors. This time constant is comparable to thin film photoluminescence data, where recombination of excitons is observed, and the theoretical hopping time expected for these particles. Briefly, the photoluminescence decay constants for CdSe/I⁻/Zn4APc films were determined to be 3.1 ns and faster (depending on film thickness, as reported elsewhere)[1] and the expected hopping time $\tau$ is 2.1 ns, calculated using $\tau = \frac{ed^2}{6\mu kT}$,[36] with the quantum dot diameter ($d$) and the mobility ($\mu$), taken here as $10^{-3}$ cm²/Vs.[1] We therefore attribute the first exponential signal decay to recombination and initial trapping of charge carriers.

Furthermore, we observe no field dependence of the FWHM for all impulse responses (cf. Figure SI3a for an example of µm polyimide devices), as opposed to Punke *et al.*[37] This shows that there is no transit time limit in the devices as faster charge carrier extraction and, thus, faster transit times would be expected for higher fields.

**Steady State.** We therefore proceed with a discussion of the steady-state data in Figure 5 to determine whether geometry-induced speed limitations are present. We observe no field dependence of the rise or fall times (10 – 90 %) in the µm devices which is a strong hint for an RC limitation as both, diffusion and drift of charge carriers, should be accelerated with increasing electric field. This implies that the measured rise times of several hundreds of µs up to several ms are the result of capacitive (dis-)charging of the device and not an intrinsic material property. This assumption is supported by the fact that the theoretically expected drift time is significantly shorter. According to $\tau_{drift} = \frac{d^2}{\mu V}$, with $\mu = 10^{-3}$ cm²/Vs[1], one expects drift times of 6.3 µs, 25 µs, 100 µs, 400 µs for the four µm channel lengths (2.5 µm, 5 µm, 10 µm, 20 µm, respectively) with 10 V bias applied.

We derive further proof for an RC limitation of the devices from the impedance spectroscopy measurements in Figure 6a. Although the capacitances of 1 – 10 pF are small, the RC product becomes significant due to the high film resistance - even under laser illumination. We note excellent agreement between the calculated RC time constant and the measured rise time if the resistance under illumination, and not the dark resistance, is used in the calculation of the RC time, see Figure 6b. This can be explained by the fact that the resistance adjusts very quickly under illumination and thus this $R_{illum}C$ time is closer to the actual time constant present at any time of the measurement.

Polyimide-based µm photodetectors are indistinguishable from glass-based µm devices under steady-state conditions. We explain this with a saturation of the trap states due to long illumination times, resulting in a comparable decay behavior. For nm glass detectors, the rise is as fast as 24 µs, but the RC limitation prevails.

**CONCLUSION**

We have investigated the speed of the photoelectrical response of CdSe/I$^-$/Zn4APc quantum dot thin films on polyimide and glass substrates. With a 3 dB bandwidth of 85 kHz, photodetectors of the same quantum dots on polyimide substrates were significantly faster than on glass with 67 kHz. While previous attempts to fabricate fast quantum dot-based photodetectors focused mainly on tuning of the active layer, the electrode material, and the electrode geometry, our results highlight the equally important role of the substrate in this respect. Moreover, we have shown that the major challenge toward faster quantum dot photodetectors is the high photoresistance, which leads to RC-limited response times.

**EXPERIMENTAL METHODS**

**Chemicals.** Acetone: Acros Organics, 99.8 %; ammonium iodide: Aldrich Chemicals, 99.999 %; cadmium oxide: Sigma-Aldrich, 99.99 %; ethanol: Acros Organics, 99.5 %; hexadecylamine: Acros Organics, 90 %; N-methylformamide: Acros Organics, 99 %; 1-octadecene: Acros Organics, 90 %; oleic acid: Sigma-Aldrich, 90 %; selenium: Aldrich Chemicals, 99.999 %; trioctylphosphine: abcr, 90 %; trioctylphosphine oxide: Sigma-Aldrich, 99 %.

**Particle Synthesis**. CdSe quantum dots were synthesized using a procedure of Sayevich et al.[38] In short, 176.0 mg CdO, 8.0 g trioctylphosphine oxide, 8.0 g hexadecylamine, 2.2 mL oleic acid, and 45.8 mL 1-octadecene were kept in a tree-neck flask under vacuum (~$10^{-3}$ mbar) for 2 h before heating to 300 °C under nitrogen to form a clear solution. The temperature was then reduced and held at 275 °C for 30 min. Afterwards, a solution of 126.0 mg Se in 4 mL trioctylphosphine, 4 mL trioctylphosphine, and 4 mL 1-octadecene was injected into the reaction mixture, the temperature was increased to 280 °C and kept at this temperature for 45 min. The reaction was quenched by a sudden cooling with cold water. The purification of the quantum dots was done by precipitating the particles twice with ethanol and twice with acetone, centrifuging after each precipitation and redispersing them in hexane. For the ligand exchange, 300 µL of a 1 M solution of NH$_4$I in N-methylformamide, 2.7 mL acetone, and 0.84 mL of a 60.5 mg/mL CdSe solution in hexane were stirred over-night, centrifuged, washed with acetone, and redispersed in 400 µL N-methylformamide.

**Device Preparation**. Commercially available bottom-gate bottom-contact Si-FET substrates (n-doped silicon, $n = 3 \cdot 10^{17}$ cm$^{-3}$) with 230 nm or 770 nm thermal oxide and gold interdigitated electrodes of 10 mm width (thickness: 30 nm) and varying channel length (2.5, 5, 10, 20 µm) of the Fraunhofer Institute for Photonic Microsystems, Dresden, Germany were used. To compare the substrates, the same electrode geometry was photolithographically prepared on 0.125 mm polyimide foil (DuPont™ Kapton® HN) and glass slides (Duran Wheaton Kimble; 76×26 mm; soda-lime glass). To this end, glass slides were coated with hexamethyldisilazane upon spincoating maP-1215 (micro resist technology) photoresist (3000 rpm; 30 s) on both substrates. The as-spinned film was soft-baked on the hot plate (100 °C) for 60 s. Optical lithography was performed with a Karl Süss Mask Aligner MA/BA6. After development in maD-331/S (micro resist technology), the substrates were metallized with 3 nm titanium and 20 nm gold in a PLS570 evaporator. The lift-off was done in acetone in an ultrasonic bath for two minutes. For the channel lengths below 2.5 µm, electron beam lithography was performed at a JEOL JSM-6500F SEM with a Xenos PG2 pattern generator. For this, an approx. 90 nm thick PMMA 2041 layer (2.5 % dissolved in MIBK) was spincoated onto a hexamethyldisilazane functionalized glass substrate (2600 rpm; 6 s; acceleration 6 s followed by 5000 rpm; 60 s; acceleration 6 s) and dried for 5 min at 150 °C. To enable conduction during exposure, a conductive polymer, Electra-92 (Allresist), was spincoated onto the resist layer (4000 rpm; 40 s; acceleration 6 s) and dried 5 min at 90 °C. For a better adhesion of these two layers, the PMMA coated substrate was activated in a Reactive Ion Etching process (100 % $O_2$; 20 s; 250 W; 250 mTorr). The electrode fingers were written with a dose of 0.3 nC/cm and were developed in a 3:1 mixture isopropanol to MIBK for 12s.

Thin film preparation was then performed in a $N_2$ filled glovebox. Typical film preparation was performed as described elsewhere.[1] In short, 4.5 nm CdSe/I⁻ QDs in N-methylformamide (~ 45 µL, 88 µM) and ~ 65 µL of a saturated Zn4APc solution in DMSO were deposited onto the substrate to be investigated and left to form a CdSe/I⁻/Zn4APc film over-night. Afterwards, remaining solvent was spun-off the substrate and the film was washed with acetonitrile to remove unbound Zn4APc. The film was finally annealed at 190 °C for 30 min. The films were transferred under $N_2$ into the probe station via a home-built transfer arm to avoid contact with air.

**Electrical Measurements**. Electrical measurements were performed at room temperature in a Lake Shore Cryotronics probe station CRX-6.5K under a pressure of $5 \cdot 10^{-6}$ mbar with a Keithley 2634B System Source Meter and the samples were contacted with tungsten two-point probes.

**Transient Photocurrent**. Photocurrent measurements were performed in the probe station at room temperature and under vacuum. Steady state photocurrent was measured using square pulse illumination of the devices. A fast switchable laser driver (FSL500, PicoQuant) with a laser rise time of < 0.5 ns together with a 635 nm laser diode with an output power of ≤ 12 mW was operated at 10 / 100 Hz, externally triggered with a Hewlett Packard 33120A arbitrary waveform generator. Impulse measurements were carried out using a picosecond pulsed laser driver (Taiko PDL M1, PicoQuant) together with laser heads for 636 nm and 779 nm illumination with pulse width of < 500 ps. Repetition rates of 100 kHz and 1 MHz were selected with average output powers of 22 µW and 220 µW, respectively, i.e., a pulse energy of 0.22 nJ. 1 MHz data was used for the exponential fitting of the first decay to ensure a good resolution of the measurement (limited by the time-resolution of our lock-in amplifier with approx. 1.7 ns), whereas the 100 kHz data was used both for the investigation of the power law dependence of the tail and the bandwidth calculation to ensure that the tail drops into noise prior to further excitation. We calculated the bandwidths of the photodetectors via the power spectrum $P(\omega)$ using the fast Fourier transformation FFT of the impulse response $f(t)$: $P(\omega) = |\text{FFT}(f(t))|^2$.[39] In order to determine the 3 dB bandwidth, we had to apply zero-padding, and thus converted the 100 kHz data into quasi 25 kHz measurements. The given, theoretical laser powers were further reduced by scattering, inefficient coupling into the optical fiber, decollimation of the beam, etc. The laser spot was not focused in order to illuminate the whole electrode area and to avoid damage to the organic dye. The electrodes were contacted using 50 Ω matched tungsten probes and 40 GHz coaxial cables chosen as short as possible. The current was preamplified with a FEMTO DHPCA-100 current amplifier and measured with a Zurich Instruments UHFLI lock-in amplifier with Boxcar Averager Function, which averages the signal from 2 G samples. The signals were background-corrected. The time resolution of the setup was limited to 600 MHz because of the signal input limitation of the lock-in amplifier.

**Impedance Spectroscopy**. Impedance Spectroscopy was carried out using a CH Instruments Electrochemical Analyzer/Workstation Model 760E. Both uncoated electrodes on polyimide and CdSe/I⁻/Zn4APc thin film on top of Au electrodes on polyimide/glass were checked. One of the electrodes was connected to the working electrode of the potentiostat, the other device electrode was connected to both reference electrode and counter electrode of the potentiostat. Typical measurements were performed from $10^5$ Hz to $10^{-3}(10^{-1})$ Hz with a measurement amplitude of 500 mV (resulting in P < 0.5 µW to prevent the sample from heating during measurement). Above 100 Hz, FT has been selected as measurement mode, below, single frequency mode is the potentiostat preselected mode. The sensitivity scale setting was set to automatic and 12 points/decade frequency were measured with 1 cycle per decade. The spectra were fitted assuming a parallel circuit consisting of a constant phase element $(Y_0, n)$ and a resistor $(R)$, using the impedance package for python.[40] The effective capacitance $C_{eff}$ was calculated via $C_{eff} = (Y_0 \cdot R^{1-n})^{\frac{1}{n}}$ according to Hsu et al.[41]

ASSOCIATED CONTENT

**Supporting Information**. Figure SI1: Substrate characterization (µm), Figure SI2: Schematic of photocurrent measurement setup, Figure SI3: Voltage and channel length dependence of 636 nm impulse measurements, Figure SI4: Exemplary impedance data, Figure SI5: Resistance of devices - Comparison of two-point measurements with impedance measurements, Figure SI6: Substrate characterization II (nm), Figure SI7: Time domain figure of Figure 8a, Figure SI8: Particle characterization

AUTHOR INFORMATION

**Corresponding Author**

*To whom correspondence should be addressed: marcus.scheele@uni-tuebingen.de

**Author Contributions**

The manuscript was written through contributions of all authors. All authors have given approval to the final version of the manuscript.

ACKNOWLEDGMENT

Financial support of this work has been provided by the European Research Council (ERC) under the European Union's Horizon 2020 research and innovation program (grant agreement No 802822. We thank Kai Braun for his help in constructing the transient photocurrent setup, Alexandru Oprea for discussions regarding the impedance spectroscopy data, and Elke Nadler for electron microscopy imaging.

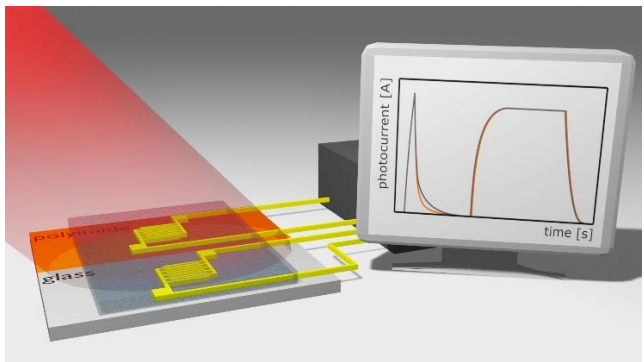

Table of content graphic.

# Supporting Information

# Substrate Effects on the Bandwidth of CdSe Quantum Dot Photodetectors

*Christine Schedel[1], Fabian Strauß[1], Krishan Kumar[1], Andre Maier[1], Kai M. Wurst[1], Patrick Michel[1], Marcus Scheele[1,2],\**

[1]Institute for Physical and Theoretical Chemistry, University of Tübingen, 72076 Tübingen, Germany.

[2]Center for Light-Matter Interaction, Sensors and Analytics LISA+, University of Tübingen, 72076 Tübingen, Germany

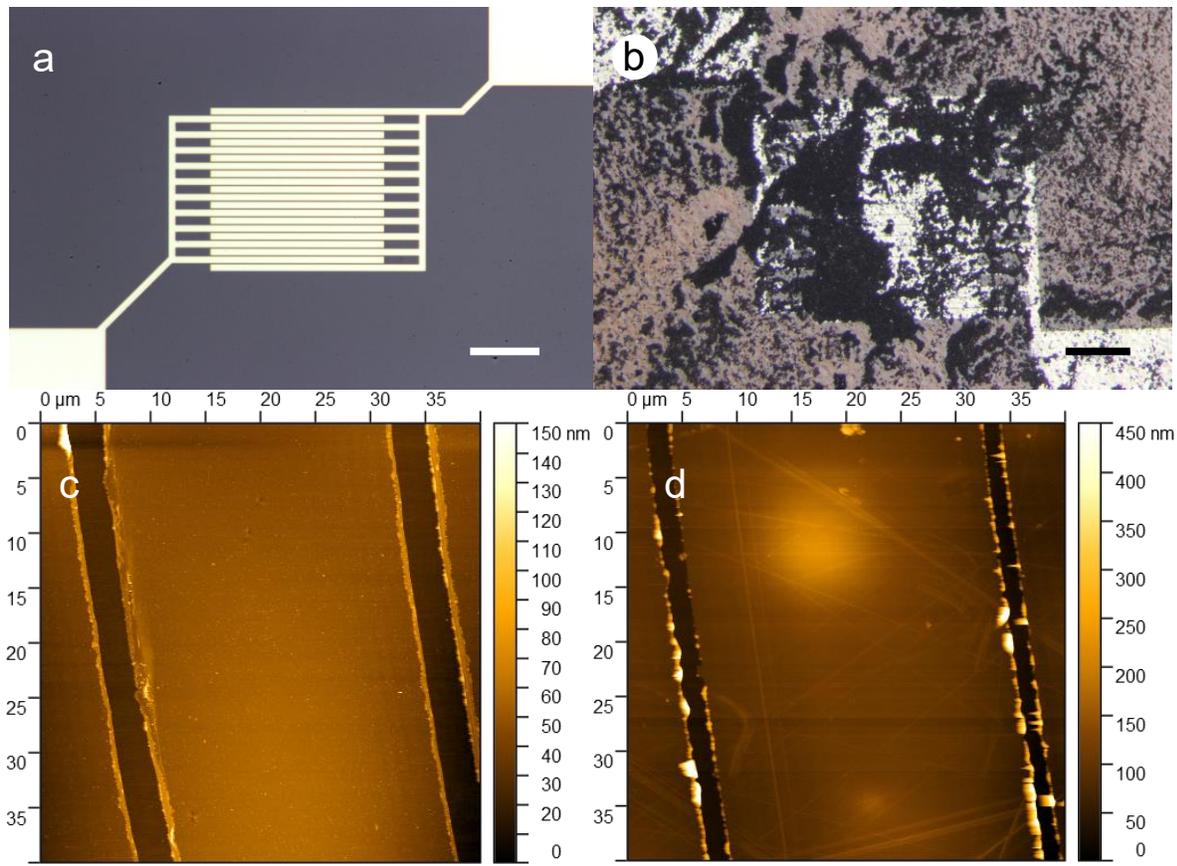

Fig. SI1: Substrate Characterization. a) First, µm-spaced electrode geometry investigated: 2.5 µm / 5 µm / 10 µm / 20 µm × 10 mm gold interdigitated electrodes, light microscope image of a 2.5 µm channel length interdigitated electrode geometry. Scale: 200 µm. b) Exemplary light microscope image of an inhomogeneous CdSe/I⁻/Zn4APc film on polyimide foil. Scale: 200 µm. AFM characterization of 2.5 µm gap electrodes on c) glass and d) polyimide, showing slightly different actual electrode gaps of approx. 3 µm and 2 µm, respectively.

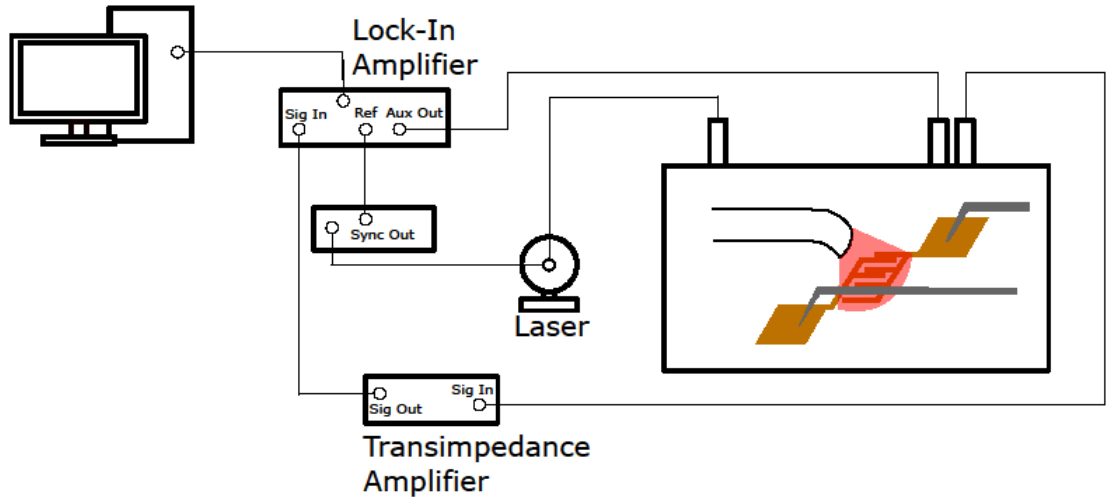

Fig. SI2: Schematic of the photocurrent measurement setup.

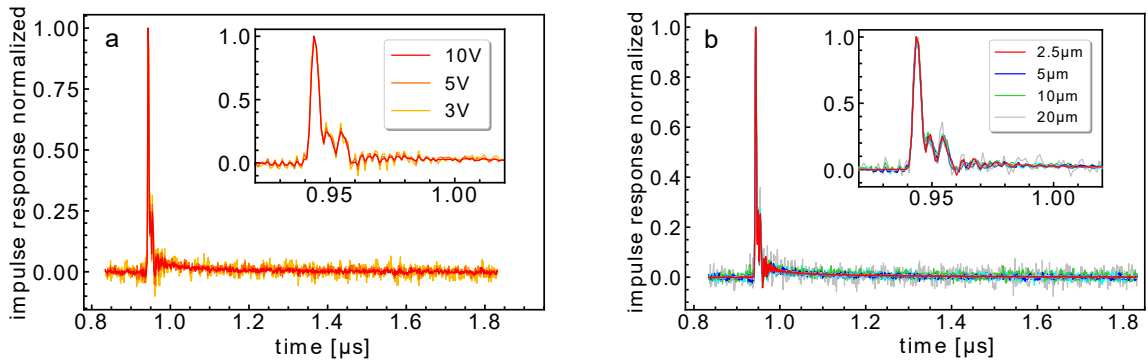

Fig. SI3: 636 nm impulse responses (1 MHz) of CdSe/I$^-$/Zn4APc on polyimide a) of one typical 5 µm channel length device with different applied electric fields, b) for different channel lengths with 10 V bias applied. In both cases, no difference in the FWHM of the photocurrent response is visible, demonstrating that there is neither a voltage nor a channel length dependence. The same is true for glass substrates and for 779 nm impulse illumination of the thin films on both substrate materials, where the photocurrent response is reduced by 50 – 80 % compared to the 636 nm response.

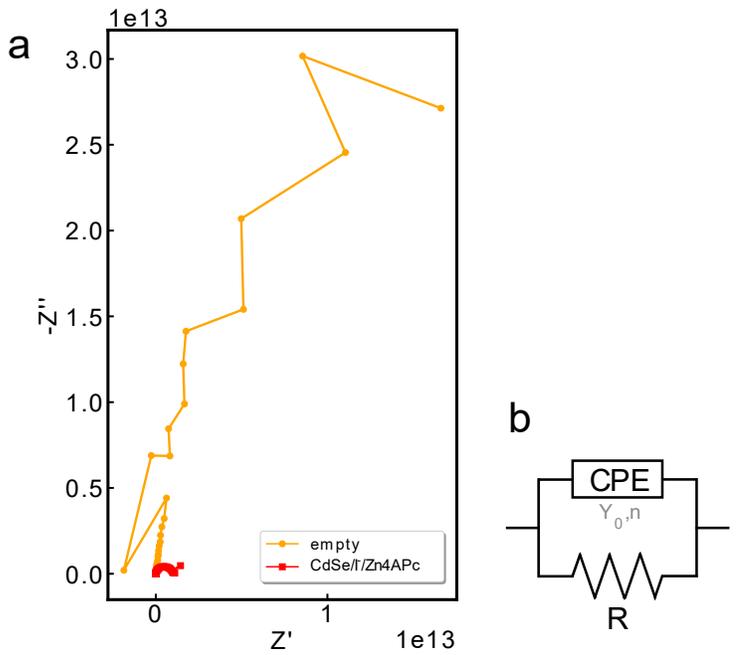

Fig SI4: Example of impedance spectra: a) Uncoated (orange) vs CdSe/I⁻/Zn4APc coated (red) gold electrodes on polyimide substrates of a 2.5 µm x 10 mm contact. b) Equivalent circuit consisting of a constant phase element $(Y_0, n)$ and a resistor $(R)$. c) Exemplary data and fit of the CdSe/I⁻/Zn4APc thin film on polyimide shown in (a), assuming the equivalent circuit shown in b.

$R = 9.0 \cdot 10^{11} \, \Omega, Y_0 = 2.73 \cdot 10^{-12} \frac{s^n}{\Omega}, n = 0.94 \rightarrow C_{eff} = 2.89 \cdot 10^{-12}$ F.

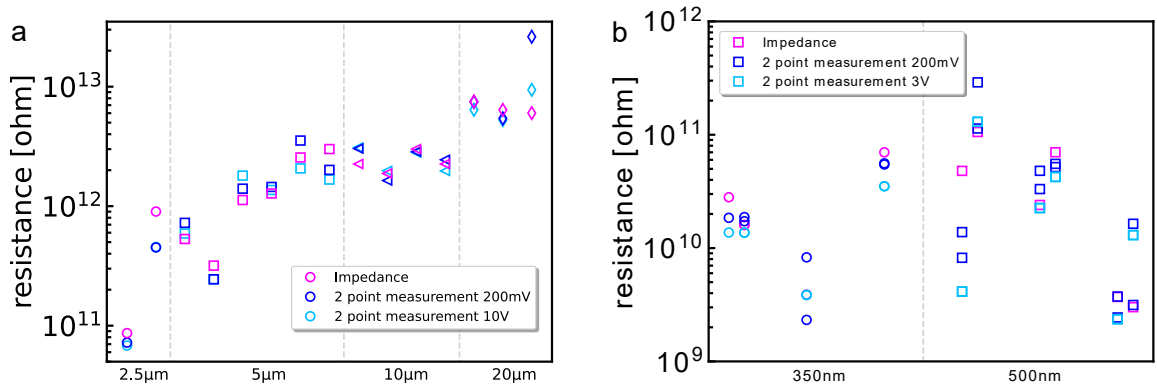

Fig SI5: Resistance of devices. Comparison of two-point measurements (blue) with impedance measurements (magenta), x-axis: electrode gap, contacts listed. a) Resistance of µm-spaced electrode devices on polyimide. b) Resistance of nm glass devices – new electrode geometry characterized in Figure SI6.

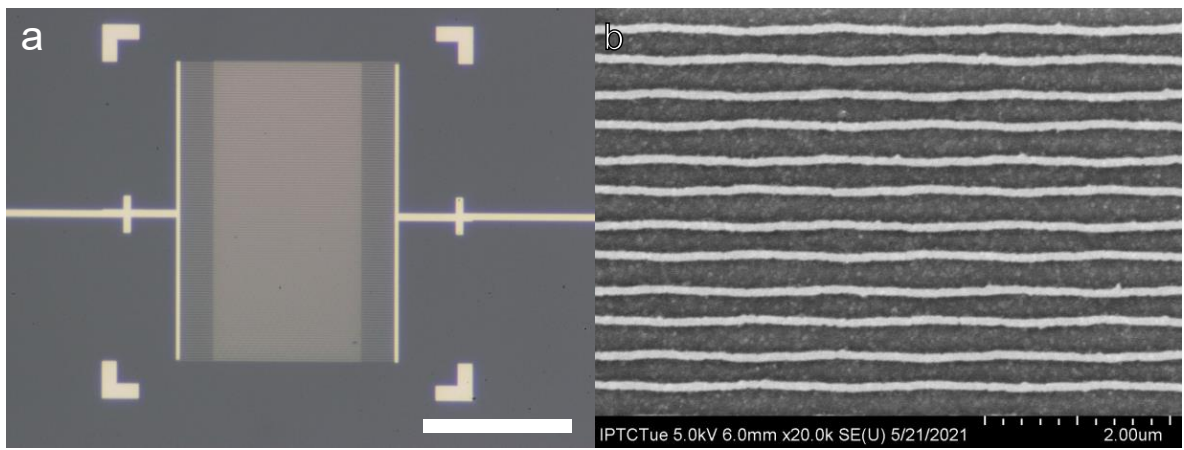

Fig. SI6: Substrate Characterization of the second, nm-sized electrode geometry with 350 nm / 500 nm × 10 mm gold interdigitated electrodes. a) Light microscope image of a 350 nm channel length device on glass. Scale: 40 µm. b) SEM characterization of the 350 nm gap electrodes on glass.

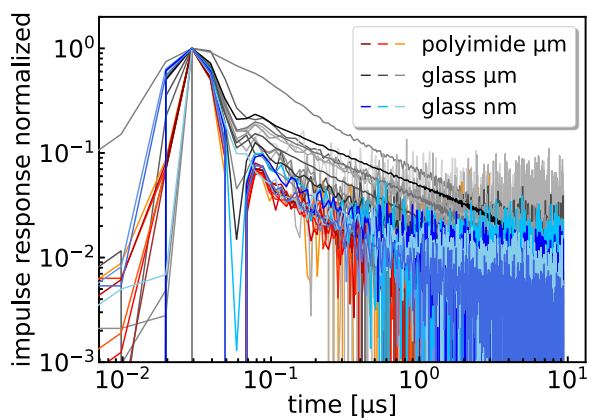

Fig. SI7: Normalized impulse response of µm- vs nm-spaced photodetector devices. Corresponding time regime figure for Figure 8a.

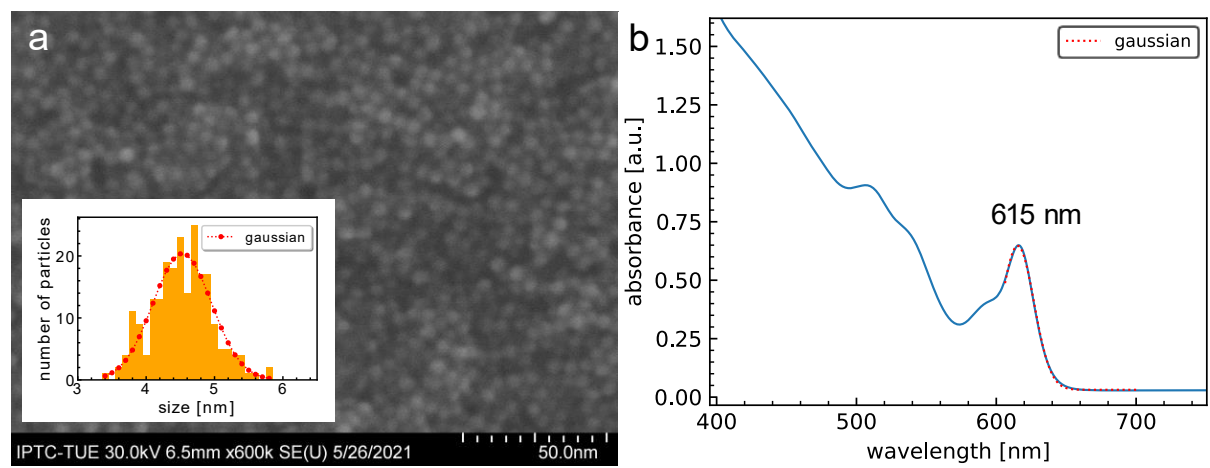

Fig. SI8: Particle Characterization. a) SEM image of CdSe/I⁻ particles used for thin film preparation, showing a size of approximately 4.5 ± 0.4 nm. b) UV/Vis spectrum of CdSe quantum dots with native ligand shell in hexane, indicating 5.3 nm particles, according to Yu et al.[1]